# Photon return on-sky test for "by-product" Rayleigh plume of sodium laser guide star

## Jian Huang[1], Lu Feng[2], Kai Jin[3], Min Li[3], Gongchang Wang[4,*]

1. College of Mechanical Engineering, Chongqing Key Laboratory of Manufacturing Equipment Mechanism Design and Control, Chongqing Technology and Business University, Chongqing, 400067, China
2. CAS Key Laboratory of Optical Astronomy, National Astronomical Observatories, Chinese Academy of Sciences, Beijing, 100101, China
3. Institute of Optics and Electronics, Chinese Academy of Sciences, Chengdu, Sichuan, 610209, China
4. Xi'an Satellite Control Center, Xi'an, 710043, China

*wh-99@126.com(Gongchang Wang)

**Abstract**: The sodium laser guide star (sLGS) adaptive optics (AO) system has become an essential component for large ground-based optical/infrared telescope. The performance of the AO system can be significantly hindered if the on-sky spot profile of its sLGS is degraded by the turbulence along the laser launching uplink path. One effective method to overcome this problem is to perform real-time "pre-correction" on the laser before it is launched onto the sky to counter this turbulence effect especially for the lower altitude where turbulence is stronger. The "by-product" Rayleigh backscatter generated by large molecule and aerosols at low altitude when projecting sLGS is a perfect candidate for real time detection of low altitude turbulence. It is therefore important to evaluate whether this "by-product" of a sodium laser guide star could achieve suitable performance for wavefront sensing. In this paper, we attempt to answer the question regarding to the achievable photon return of such Rayleigh backscattered plume. A comparison between results from our field test and theoretical model of MSISE-90 was presented. We showed that the greatest differences is less than 20% which hints the applicability of using the defocused Rayleigh plume as wavefront sensing signal for different turbulence strengths. The result also showed that in case of strong (Fried parameter ($r_0$) of 5 *cm*) or moderate ($r_0$ of 10 *cm*) turbulence, the optimum center for range gating is 9 *km*, and the corresponding value of range gate depths are 3 *km*, 1.1 *km*, 0.56 *km* and 0.28 *km* when subaperture sizes are 3 *cm*, 5 *cm*, 7 *cm* and 10 *cm*, with the maximum pulse repetition frequency at 1500 Hz.

## 1. Introduction

Adaptive Optics (AO) system detects the wavefront aberration induced by atmospheric turbulence and corrects aberrated wavefront in the optic train of the telescope with active optical component in real time to achieve near diffraction limited image quality for ground based large optical/infrared telescope [1]. The system requires a bright target such as bright natural stars in the proximity of the scientific objects for wavefront sensing. For a 10 meters class telescope, such a star should typically be brighter than 14 magnitude with angular offset from the object smaller than 10 arcsecs to achieve a strehl ratio of 0.4 [2]. This requirement severely limits the application of AO system with natural guide star (NGS), and astronomers later proposed to project laser onto the sky utilizing Rayleigh effect or resonance fluorescence of certain atoms, especially the sodium atom, to generated bright Laser Guide Star (LGS) spot on sky artificially for turbulence detection. The sodium LGS technique was tested to be effective by many observatories [3], and nowadays, it has been used in most existed 10-meter class telescopes, even planned to be installed in the next-generation large telescopes, such as the Thirty Meter Telescope, European Extremely Large Telescope and Giant Magellan Telescope as their first light instruments [4-5].

The sodium laser guide star technique utilizes the fact that there is a sodium layer existed at the altitude of 90-120 *km* in the atmosphere. If a laser tuned $D_{2a}$ line (589.15 *nm* in vacuum), the sodium atom will be excited, and it will produce resonance fluorescence emitted in all directions generating a bright artificial guide star which could be used for the AO system [6]. When the sodium laser is projected to the sky, it will inevitably passing through low atmosphere and leaving a significantly bright trail caused by Rayleigh backscatter due to aerosols and large molecules, such as $O_2$ and $N_2$. The brightness of this Rayleigh backscatter can be nearly three to four times brighter than the sodium LGS. In conventional sLGS AO system, the Rayleigh backscattered plume could leave spurious light tracks on the wavefront sensor (WFS) or cameras and it will affects the whole astronomical observation procedures in various aspects [7]. However, this "detrimental by-product" has its own advantage, it is much brighter than the sodium LGS, which means it could deliver more photon to the WFS than sodium LGS. Although its height might limit the detection volume of turbulence, it still might be used alongside with sodium LGS to perform wavefront sensing with innovative designs.

Because the Rayleigh backscatter is generated by the same laser beam for the sodium LGS, the atmospheric turbulence that affect the shape of sodium LGS spot, especially the low altitude atmospheric turbulence which has the strongest influence can be measured by the Rayleigh backscatter. This enables the trick which used Rayleigh backscatter's measurement to partially pre-correct the laser output beam for the sodium LGS to improve the shape of LGS, which in return would improve the overall turbulence detection and wavefront reconstruction performance of the AO system. In 2003, such an uplink wavefront correction system was reported by SOR, the system achieves a sodium LGS with a spot size nearly 50% smaller that the uncompensated one. The pre-correction of the so called uplink wavefront correction subsystem is mainly a 60 actuator AO system installed within the laser launching optics, the turbulence is detected by the 50 cm diameter laser launch telescope pointed at an NGS [8]. The Villages AO system installed on the Nickel telescope at Lick observatory also demonstrated the possibility using a pre-corrected sodium LGS for AO application [9-10]. But the wavefront sensing for the uplink pre-correction systems using NGS is limited by the availability of bright NGS. Jian Huang has analyzed the degradation of the sodium LGS spot mainly caused by the ground layer turbulence on the uplink path, their next step of work is detecting the turbulence on the uplink by the "by-product" Rayleigh LGS, and then controlling the power density of the sodium laser at sodium layer by pre-correction [11]. Similarly, at the same year, Jie Liu has found the same problem and a same solution has been proposed [12].

Whether the "by-product" Rayleigh LGS is used for sodium LGS pre-correction, the brightness of the Rayleigh plume is the most important factor to be considered. It is affected by the molecule numbers greatly. One of the most widely used theoretical model predicting molecular density is the MSISE-90 model [13]. To use it for estimating the photon return of the "by-product" Rayleigh LGS, it is necessary to validate the MSISE-90 atmospheric model with field test results. Secondly, the traditional Rayleigh LGS is focused at the center of the range gate, generally is at 8-12 *km* [3], while the "by-product" Rayleigh LGS generated by the plume is defocused in this altitude range, but focused at sodium layer. Therefor it is important to understand the impact of this defocusing effect for wavefront sensing, and help to achieve an optimized design of the WFS of the pre-correction system. In this paper, the "by-product" Rayleigh LGS is optimized for measurement error, cone effect and launching and receiving time sequencing these parts, and key parameters such as the center of range gate, range gate depth and maximum pulse repetition frequency were obtained.

In this paper, we focused on the study of photon return performance of the "by-product" Rayleigh plume to support our theoretical claim, and analyze differences between the "by-product" Rayleigh LGS and the traditional one. In section 2, the methodology in theoretical and the optical setup for the on-sky test were described. In section 3, experimental results for the photon return of the Rayleigh plume were presented. In section 4, key parameters for the design of the "by-product" Rayleigh LGS related to the laser or range gate were discussed. In section 5, conclusions and future works were given.

## 2. Methodology and set up of the test

The performance for wavefront sensing is evaluated by measurement errors which is inversely proportional to the "by-product" Rayleigh backscattered photons. The achievable photon return is determined by the numbers of molecules and aerosols which is related to the atmospheric model. Therefore, it is priority to verify whether the atmospheric model matches the actual situation for predicting the Rayleigh backscattering photons, then, an optimal "by-product" Rayleigh LGS can be designed by the atmospheric model.

### 2.1 Theory of Rayleigh scattering

The number of received photons from Rayleigh scattering depends on the backscattering cross-section of large molecules and aerosols in the air, the numbers of scatters which interacted with the laser field, and the power of the laser. The Lidar equation defines the detected photon numbers for a laser beam propagating through the atmosphere [3]:

$$N_R = \frac{Q \cdot \sigma_R^B \cdot T_{sys} \cdot Q_{ECCD} \cdot \lambda \cdot \rho(H_{LGS}) \cdot \Delta H_{LGS} \cdot A}{4\pi hc \cdot H_{LGS}^2 \cdot f_L} \tag{1}$$

Where $Q$ is the power of the laser, $W$; $\sigma_R^B$ is the effective backscattering cross-section, $m^2$; $T_{sys}$ is the transmission of optical components in transmit and receive paths; $Q_{ECCD}$ is the quantum efficiency of photon detector; $\lambda$ is the optical wavelength, $m$; $\rho(H_{LGS})$ is the molecule number density at the altitude of $H_{LGS}$, $atoms/m^3$; $\Delta H_{LGS}$ is the range gate length, $m$; $A$ is the area of receiving telescope, $m^2$; $h$ is the Planck's constant, $6.626 \times 10^{-34} Js$; $c$ is the velocity of light, $3 \times 10^8 m/s$; $H_{LGS}$ is the center of the range gate, $m$; $f_L$ is the temporal pulse repetition frequency of the sodium laser, Hz.

From the equation, we can observe that $\Delta H_{LGS}$ and $H_{LGS}$ are determined by the design of laser and the receiving system, $\sigma_R^B$ and $\rho(H_{LGS})$ are determined by the atmosphere, thus $\sigma_R^B$ and $\rho(H_{LGS})$ should be confirmed first.

Measures. R. M gives the Rayleigh backscattering cross-section as [3]:

$$\sigma_R^B(\lambda) = 5.45 \cdot \left[\frac{550}{\lambda(nm)}\right]^4 \times 10^{-32} m^2 \quad (2)$$

Thus, for sodium laser, the Rayleigh backscattering cross-section is $4 \times 10^{-32} m^2$.

Due to the density of $O_2$ and $N_2$ is much higher than other gases, so only these two molecules were considered in our discussion. The MSISE-90 model describes the molecule number density profile and temperature profile from 0 to 120 $km$ [13]. Our field test was conducted at Xinglong observatory, with a latitude of 40°23'47.58" N, and a longitude of 117° 34'52.50" E and the altitude is 870 $m$, the exponential fitting profile according to the measurement data can be expressed as equation (3) and shown in Fig. 1:

$$\begin{cases} \rho(H) = e^{44.6694 - 0.103 \cdot H} & 0 \leq H \leq 12 \ km \\ \rho(H) = e^{45.3586 - 0.162 \cdot H} & 12 \leq H \leq 35 \ km \\ \rho(H) = e^{44.1546 - 0.129 \cdot H} & 35 \leq H \leq 60 \ km \end{cases} \quad (3)$$

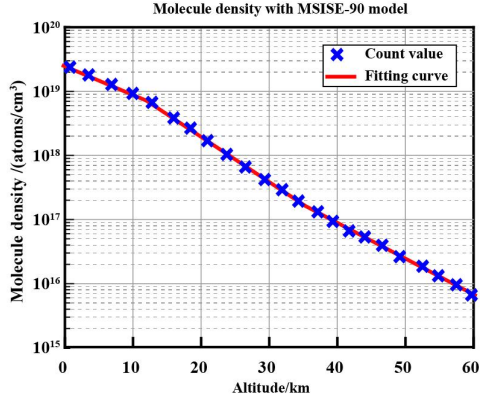

Fig. 1. Atmosphere molecule density with MSISE-90 mode, the blue cross symbols represent the measured data, the red line represent the fitting profile. The molecule density is exponentially decreased with the altitude increased.

*2.2 Set up of the test*

The experimental platform for photon return of Rayleigh plume was set up by Institute of Optics and Electronics, Chinese Academy of Science at Xinglong observatory, a sodium laser launch system pointing at zenith was used to generate a sodium laser guide star, a receiving telescope with a camera attached was used to observe the sodium laser guide star and the rayleigh scattering. The layout of the test was shown in Fig. 2. a), the distance between the sodium laser launch system and the receiving telescope is 25 $m$. The launching and receiving telescopes were setup as such so that the Rayleigh plume and the sodium LGS can be separated in the field of camera conveniently.

The sodium laser launch system is consisted of a sodium laser, beam transfer optics and laser launch telescope (LLT) pointing at zenith, the schematic diagram is shown in Fig. 2. b).

1) Sodium laser

The Quasi-Continuous Wave (QCW) pulsed solid state sodium laser was designed and fabricated by Technical Institute of Physics and Chemistry (TIPC), Chinese Academy of Sciences. The laser utilized a so-called MOPA (Master Oscillator Power Amplifiers) design for its 1064 $nm$ and 1319 $nm$ lasers and combined the two into the final 589 $nm$ beam by employing a sum frequency generation stage with Lithium triborate crystal. The pulse repetition frequency (PRF) is adjustable from 500 to 1000 Hz [14-15]. The main parameters of the laser were shown in table 1 and the picture of the laser was shown in Fig. 3. a).

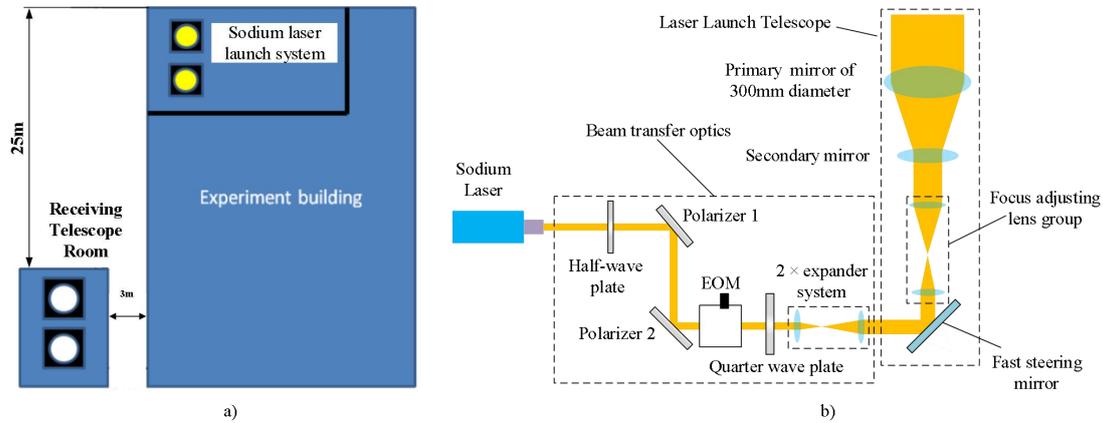

Fig. 2. a) The layout of the experiment, the distance between the laser room and the receiving telescope is 25 *m*, so that the Rayleigh plume and the sodium LGS can be separated in space conveniently. b) schematic diagram of the laser launch system, the power of the sodium laser was attenuated by HWP and two polarizers, the EOM was used for re-pumping, the QWP was used to change the polarization state of the pump laser to circular. After expanding the diameter of the laser to a proper size, the laser beam was stabilized by a fast steering mirror and focused at the sodium layer by the 300 *mm* diameter LLT.

2) Beam transfer optics and laser launch telescope

The beam transfer optics was shown in Fig. 3. b). The output linear polarized laser passing through the half-wave plate(HWP) and two polarizers, the power of the laser can be attuned by changing the rotation angle of the HWP. The Electro-Optic Modulator (EOM) were used to modulate 10% of total laser output power from the original $D_{2a}$ line to the sodium $D_{2b}$ line for $D_{2b}$ repumping [16]. A Quarter Wave Plate (QWP) was placed after the EOM to adjust the final polarization state of the pump laser to circular to enhance the photon return of the sodium LGS [17]. A 2× expanding lens group with focal length of 100 *mm* and 300 *mm* was placed before the laser entering the LLT which was used to expands the laser beam to fill the primary mirror of the LLT.

The diameter of the LLT was 300 *mm* and the light path was shown in Fig. 3. c). The LLT was mainly consisted with three parts: the beam expanding system consisted with a refractive aspherical primary mirror and expanding secondary mirror, the imaging and the relay system. Flexures of the LLT could induce pointing error for the position of the LGS in the sky. The LLT can be used as an imaging telescope for the star, with a fast steering mirror in the relay light path to compensate for any mechanical pointing error. The position of the sodium LGS can be stabilized in real time and all the subapertures can be illuminated. The diameter of the laser was expanded to 240 $mm@1/e^2$ at primary mirror, then was projected by the LLT and the sodium laser was focused on the sodium layer by adjusting the lens #1 and lens #2 at different altitude angles. The laser diagnostic system measures the laser power, the polarization of the pump laser and the beam quality. The main parameters of the laser launch system were shown in table 1 as described in [18].

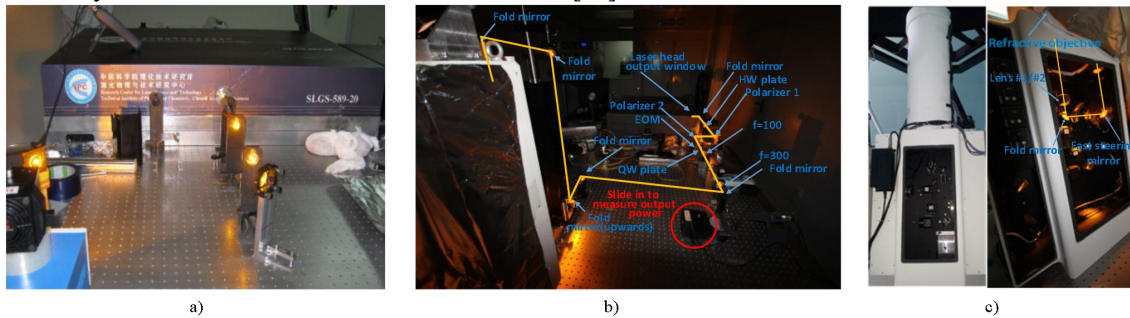

Fig. 3. a) The sodium laser designed and fabricated by Technical Institute of Physics and Chemistry (TIPC), Chinese Academy of Sciences. b) The beam transfer optics in our experiment. The HWP, polarizers and an EOM was used to modulate10% of total laser power for $D_{2b}$ repumping. The QWP was placed after the EOM to adjust the final polarization state of the pump laser to circular. The expanding lens group with focal length of 100 *mm* and 300 *mm* was used to expand the diameter of the laser to a proper size. c) The 300 *mm*-diameter LLT used to project the sodium laser and the laser launching optical path in the LLT, the laser from the beam transfer optics was stabilized by a fast steering mirror. The laser was focused at the center of sodium layer by the primary mirror and adjustment of the lens #1 and lens #2 at different altitude angles accordingly.

Table 1. Main subsystems and their nominal values for the experiment

| Laser | | LLT | | CDK Receiving telescope | |
|---|---|---|---|---|---|
| Parameter | value | Parameter | value | Parameter | value |
| Power | 20 $W$ | Out diameter | 300 $mm$ | Central obstruction | 23.5% |
| PRF | 500 Hz | Throughput | 0.815 | Out diameter | 14 inches |
| Pulse length | 100 $\mu s$ | Focusing | 90~210 $km$ | Focal length | 2563 $mm$ |
| Beam quality | $M^2$~1.25 | Beam diameter | 240 $mm$@$1/e^2$ | Pixel count | 512×512 |
| Wavelength | 589 $nm$ | Fiele of view | ±2.5' | Pixel scale | 1.93''/pixel |

3) Receiving telescope

A Planewave Corrected Dall-Kirkham (CDK) telescope was used as the receiving telescope (Fig. 4). During observation, the telescope was always in lock position with sodium LGS in the center of the field having the Rayleigh plume as well in the field of view. A Princeton Instrument Versarray CCD was installed as backend. A standard Astrodon Johnson UBVRI V band filter was inserted before the camera. The main parameters of the receiving telescope were listed in table 1.

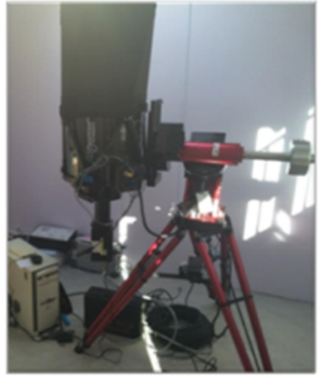

Fig. 4. The Planewave Corrected Dall-Kirkham (CDK) telescope was used to observe the sodium LGS and the Rayleigh plume.

## 3. Experiment results

In the experiment, the sodium LGS and the Rayleigh plume were both obtained, as shown in Fig. 5. a). Image field was rotated in such a way that the Rayleigh scattering is vertical, 10 images with 1s exposure were gathered as a group, a region horizontally nearby the sodium LGS was selected and its median was used as sky background.

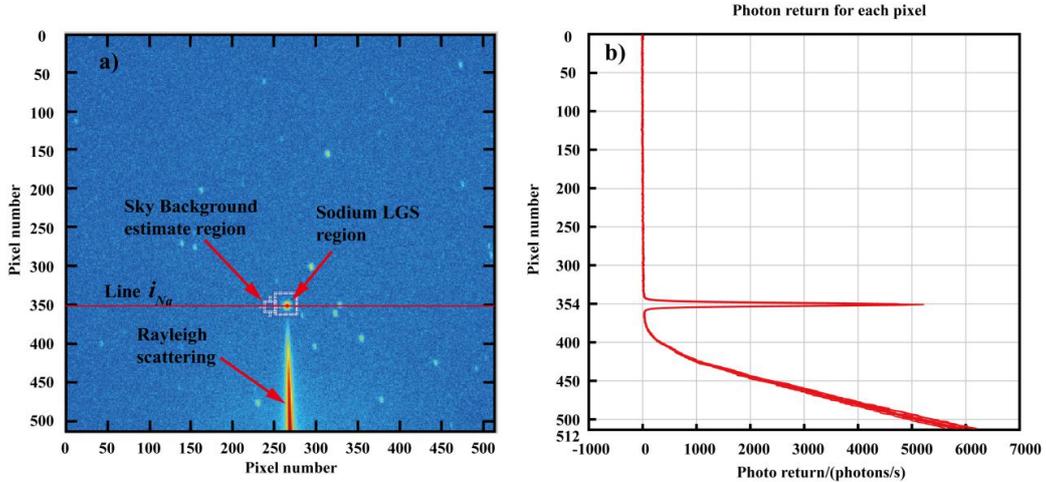

Fig. 5. a) Example of the sodium LGS and the plume of the Rayleigh scattering. A region horizontally nearby the sodium LGS and used its median as sky-background, the center of the sodium LGS is marked in the line $i_{Na}$. b) Ten frames of the photon return profile.

The photon return profile along the vertical axis of pixels in the picture is shown in Fig. 5. b). Ten frames of the measurements were shown in the figure. The photon return dominated by Rayleigh scattering are the pixels fall to the 380th, which is proportional to the molecule density in the atmosphere. There is a strong peak around 360th pixel which is caused by the sodium LGS.

In order to verify the photon return's variation with altitude, the altitude and the altitude range represented by each pixel should be determined. The center of the sodium LGS is in line $i_{Na}$. The altitude angle of the sodium LGS $\theta_{Na}(°)$ is defined as:

$$\theta_{Na} = (512 - i_{Na}) \times 1.93" \tag{4}$$

In reality, the height of the sodium LGS is:

$$\theta'_{Na} = \arctan\frac{90000}{25} = 89.98408451° \tag{5}$$

The pixel in line 512 correspond to the lowest altitude:

$$\theta_{512} = \theta'_{Na} - \theta_{Na} \tag{6}$$

So the height of pixel in line $i$ (unit is $m$) was denoted as $h_i$, which can be expressed as:

$$h_i = 25 \times \tan(1.93" \times (512 - i) + \theta_{512}) \tag{7}$$

And the altitude range represented by certain pixel (unit is $m$) in line $i$ was denoted as $\Delta h_i$, which can be expressed as:

$$\Delta h_i = 25 \times \tan(\theta_{512} + (512-i) \times 1.93 + \frac{1.93}{2}) - 25 \times \tan(\theta_{512} + (512-i) \times 1.93 - \frac{1.93}{2}) \tag{8}$$

Thus, the Rayleigh scattering versus the altitudes can be extracted from the images. The theoretical (from equation (1)) and experimental results were shown in Fig. 6. a), at the same time, 7 groups of measurement were conducted, the average difference (in equation (9)) and peak difference (in equation (10)) were shown in Fig. 6. b).

$$\Delta = \frac{\int |N^h_{theory} - N^h_{experiment}| dh}{\int N^h_{theory} dh \cdot \int dh} \tag{9}$$

$$\Delta_{max} = \max_{h \in (14km, 53km)} \left\{ \frac{(N^h_{theory} - N^h_{experiment})}{N^h_{theory}} \right\} \tag{10}$$

where $N^h_{theory}$ represents the photons of Rayleigh scattering at the altitude of $h$ from theory calculation, $N^h_{experiment}$ represents the photons of Rayleigh scattering at the altitude of $h$ from our experiment, $\Delta$ represents the average difference of Rayleigh scattering along the path, $\Delta_{max}$ represents the peak different, $\max\{\bullet\}$ means the maximum value in the sequence.

From the curves, the photon return is exponentially decreased due to the variation of molecule density. The photon return from Rayleigh backscatter nearly disappears for altitudes above 40 $km$. Comparing the theoretical and experimental results, the greatest difference is nearly 20% but the average difference is less than 1%. The reason that the experimental results is higher is that a result both from Mie scattering from dust and aerosols and local irregularities in the actual molecule density profile than the model. This test results showed that the MSISE-90 model could have a good match with field measurement. Therefore, it is applicable for predicting the photon return of the "by-product" Rayleigh LGS for wavefront sensing of pre-correction.

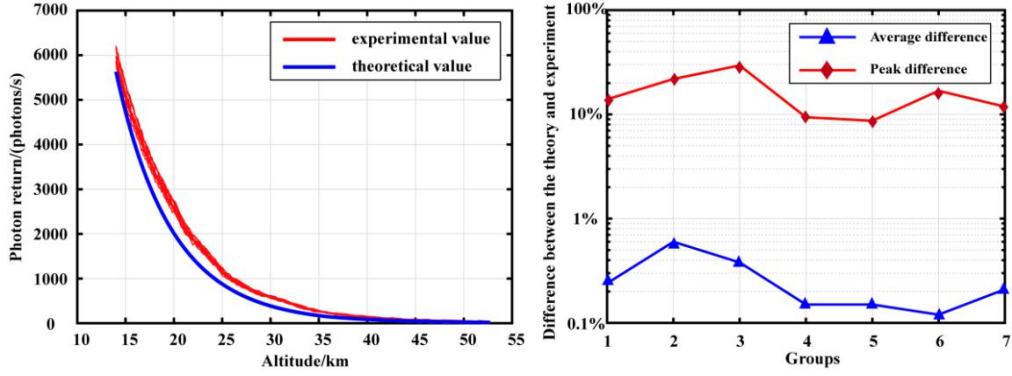

Fig. 6. a) The plume of Rayleigh scattering extracted from the experiment image (in red line) compared with the theoretical value (blue line), the photon return varies with altitude. b) the average difference and peak difference between experiment and theoretical result of 7 groups tests.

## 4. The photon flux and AO performance

The atmospheric model and theory have been verified in section 3. In this part, a "by-product" Rayleigh LGS can be designed according the atmospheric model and theory. To evaluate the performance of AO system with the LGS, the measurement error in Shack-Hartmann (S-H) WFS is used as the evaluation criteria. Due to the finite altitude of the "by-product" Rayleigh LGS, the cone effect should be discussed. The launching and receiving time sequencing related to pulse repetition frequency of the laser should also be considered because the turbulence detected by the previous pulse should be separated from the next one.

*4.1 The measurement error*

The S-H WFS measurement error $\sigma^2$ depends on the spot size (FWHM) and the signal to noise ratio (SNR) per subaperture per frame. The variance in one-axis can be expressed as [3]:

$$\sigma^2 = \left(\frac{\pi^2 K_g}{4(SNR)}\right)^2 \left(\left(\frac{3}{2}\right)^2 + \left(\frac{\theta d}{\lambda}\right)^2\right) \quad r_0 > d$$

$$= \left(\frac{\pi^2 K_g}{4(SNR)}\right)^2 \left(\left(\frac{3d}{2r_0}\right)^2 + \left(\frac{\theta d}{\lambda}\right)^2\right) \quad r_0 < d \tag{9}$$

Where $K_g$ is a factor, represented the increase in error at null, nearly 1.2~1.5. In this paper, we defined it as 1.4; $d$ is the diameter of the subaperture, $m$; $\theta$ is the FWHM of the LGS on one axis, $rad$; $\lambda$ is the wavelength of the LGS, $m$; SNR is the signal to noise ratio for each subaperture each frame.

According to equation (9), if the output power of the laser and the atmospheric model were determined, the key parameters of the Rayleigh LGS are the spot size and the photon return, which were mainly influenced by the range gate length $\Delta H_{LGS}$ and the center of the range gate $H_{LGS}$. In this part, the Rayleigh LGS was assumed to be used for pre-correction, we analyzed these key parameters for different subaperture diameter at LLT entrance pupil for 3 $cm$, 5 $cm$, 7 $cm$ and 10 $cm$.

For S-H WFS used for detecting the turbulence, the SNR is dominated by the number of backscattered photons from the Rayleigh LGS reaching the WFS. Because the brightness of the Rayleigh LGS is very high, the SNR can be easily expressed as [19]:

$$SNR \approx \sqrt{N} \tag{10}$$

Where $N$ is the average number of photons received in a S-H subaperture each frame, which is exponentially decrease when altitudes increases. In order to ensure a sufficient number of photon return, with higher $H_{LGS}$, the $\Delta H_{LGS}$ should also be larger. In practice, a 9~10 SNR for each subaperture per frame is sufficient to meet the requirement of AO system. In this paper, we assumed a fixed SNR of 10. The relationship between the $H_{LGS}$ and $\Delta H_{LGS}$ according to equation (1) with different subaperture size was shown in Fig. 8. a). From the figure, the $\Delta H_{LGS}$ exponentially increased with the $H_{LGS}$ due to the molecule number density. Usually, the range gate should be less than 3 $km$. In this case, when the subaperture size are 5 $cm$, 7 $cm$ and 10 $cm$, the center of the range gate can reach 12 $km$, but for subaperture size of 3 $cm$, the center of the range gate could not increase more than 9 $km$.

The spot size of the "by-product" Rayleigh LGS is different from the traditional Rayleigh LGS

which is focused at the center of the range gate. The spot size of the "by-product" Rayleigh LGS is the plume along the axis of the laser on the uplink path. For a sodium laser LLT, as shown in Fig. 7, the laser would be focused at the sodium layer, the "by-product" Rayleigh LGS generally would be generated at an altitude below 12 $km$, the distance between the "by-product" Rayleigh LGS and sodium LGS is larger than the Rayleigh range $Z_R$, the spot size of the "by-product" Rayleigh LGS can be determined geometrically and can be expressed as [20]:

$$D_i \approx \frac{D_{LLT}}{2 \cdot H_{LGS}} \left| 1 - \frac{H_{LGS}}{H_{Na}} \right| \qquad (11)$$

Where $D_{LLT}$ is the diameter of the LLT, $m$; $H_{Na}$ is the center altitude of the sodium LGS, which is nearly 90 $km$. The spot size at different altitudes and relevant range gate depth are shown in Fig. 8. b). However, while the laser beam approaching the focal point at sodium layer, the spot size of the sodium LGS does not approach to 0 [21].

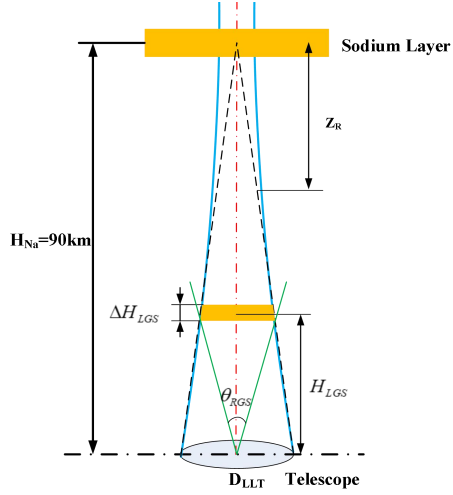

Fig. 7. The cross section of Rayleigh plume from the geometrical mode (in dashed line) and the real laser beam (in blue line).

The spot size is also related to the spot elongation induced by range gate length. It can be seen that the elongation will become larger in radial direction from the axis of the LLT. The angular size of the maximum elongation can be expressed as:

$$\alpha = \frac{D_{LLT} \cdot \Delta H_{LGS}}{2 H_{LGS}^2} \qquad (12)$$

Where α is the elongation in radius; $D_{LLT}$ is the diameter of the laser launch telescope, $m$; $\Delta H_{LGS}$ is the range gate depth of the Rayleigh LGS, $m$; $H_{LGS}$ is the center altitude of the range gate of the Rayleigh LGS, $m$. For different range gate depth and the center altitude of the range gate, as shown in Fig. 8. a), the elongated spot is shown as Fig. 8. c). The optimal angular size of the Rayleigh LGS $\alpha_{opt}$ is determined by the diffraction limit of the subaperture or the value of $r_0$ [3]:

$$\begin{aligned} \alpha_{opt} &= 1.22 \frac{\lambda}{d} \qquad d \leq r_0 \\ &= 1.22 \frac{\lambda}{r_0} \qquad d > r_0 \end{aligned} \qquad (13)$$

As shown in Fig. 8. b), the diffraction limit for subaperture size at LLT entrance pupil of 3 $cm$ (green dashed line), 5 $cm$ (blued dashed line), 7 $cm$ (orange dashed line) and 10 $cm$ (purple dashed line) are 5 arcsec, 3 arcsec, 2 arcsec, 1.5 arcsec respectively. Usually, the pre-correction system was used for strong turbulence (usually $r_0$ is nearly 5 $cm$), the lowest altitude for the center of the range gate is nearly 9 $km$.

The field of view (FOV) of the subaperture must be large enough that the elongated spot can not contaminate adjacent subapertures. According to Backer's research, an elongation larger than 1 arcsec will substantially decrease the sensitivity of wavefront sensing in the elongated direction for photon limited applications [22]. Thus the center of the range gate is below 9 $km$ for the diameter of subaperture is 3 $cm$.

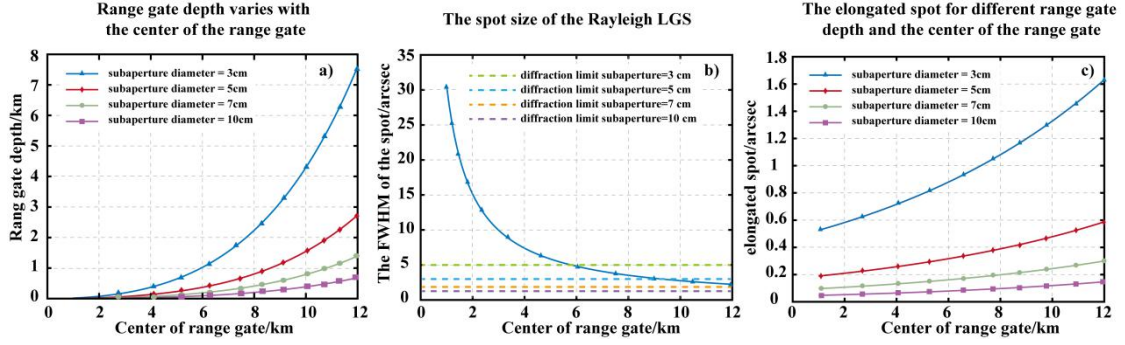

Fig. 8. a) For the SNR is 10, the range gate depth varies with the center of range gate due to the exponential decrease in molcule density. b) The spot size calculated by the geometrical mode. The diffraction limit for the subaperture with diameter of 3 *cm*, 5 *cm*, 7 *cm* and 10 *cm* were plotted with green, blue, orange and purple dashed lines respectively. c) The spot elongation for the Rayleigh plume in subaperture with the range gate depth and center of range gate as a).

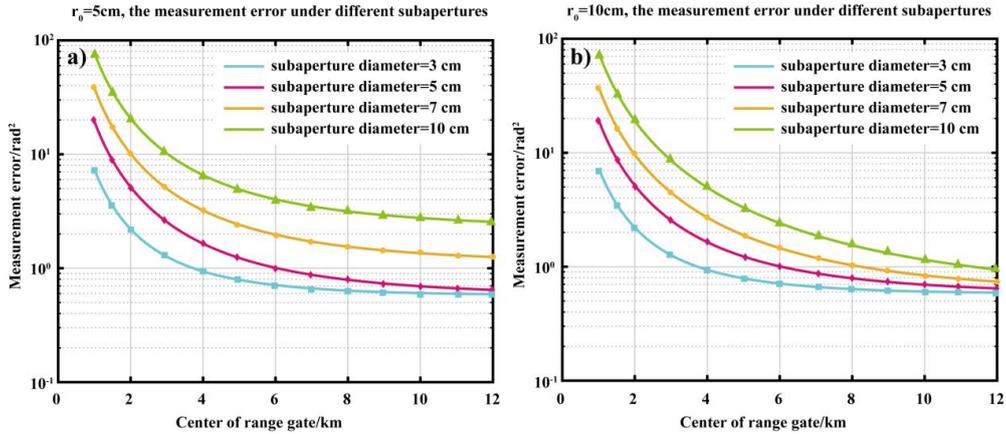

Fig. 9. The measurement error for a) $r_0$ is 5 *cm* and b) for $r_0$ is 10 *cm*, subaperture size is 3 *cm*, 5 *cm*, 7 *cm* and 10 *cm* respectively, the error is converging for the center of the range up to 9 *km*. but for the $r_0$ is 10 *cm*, the errors of subaperture size is 3 *cm* and 5 *cm* is less than that of $r_0$ is 5 *cm*.

According to the previous results, the measurement error was calculated in two directions. One is parallel to the direction of elongation, in this case, $\theta$ is equal to the size of elongated spot. For the direction perpendicular to the direction of elongation, $\theta$ is equal to the spot size. The total measurement error (the two-axis) is the sum of these two errors. Because the subaperture size is less than 10 *cm* for pre-correction with the value of $r_0$ less than 10 *cm*, the typical cases of measurement error for $r_0$ is 5 *cm* and 10 *cm* were analyzed. In Fig. 9. a), for $r_0$ equals to 5 *cm*, the errors with different subaperture size decreased when the center of range increased. However, when the center of range is higher than 9 *km*, both errors are converging. The converged value is approximately 2.5 $rad^2$ and 1.3 $rad^2$ for subaperture size of 10 *cm* and 7 *cm* respectively, 0.6 $rad^2$ for subaperture size of 5 *cm* and 3 *cm*. In Fig. 9. b), $r_0$ equals to 10 *cm*, the converged value of the error is 0.9 $rad^2$, 0.7 $rad^2$ for subaperture size of 10 *cm* and 7 *cm* respectively, 0.6 $rad^2$ for subaperture size of 5 *cm* and 3 *cm*. The center of range gate can be selected as 9 *km*, the corresponding value of range gate depth is 3 *km*, 1.1 *km*, 0.56 *km* and 0.28 *km* for subaperture size is 3 *cm*, 5 *cm*, 7 *cm* and 10 *cm*.

*4.2 The cone effect*
The turbulence above the LGS can not be detected and the turbulence below the LGS is only partially sampled. The resulting wavefront error is called the cone effect (focal anisoplanatism), which can be expressed as $\sigma_{cone}^2$ ( in $rad^2$ ):

$$\sigma_{cone}^2 = (\frac{D_{LLT}}{d_0})^{5/3} \qquad (14)$$

Where $D_{LLT}$ is the diameter of the laser launch telescope, *m*; $d_0$ can be considered as the diameter of the aperture over that the wavefront error induced by cone effect is 1 $rad^2$, which can be expressed as:

$$d_0 = \left\{ k^2 \left[ 0.057 \mu_0^+(H) + 0.500 \frac{\mu_{5/3}^-(H)}{H^{5/3}} - 0.452 \frac{\mu_2^-(H)}{H^2} \right] \right\}^{-3/5} \quad (15)$$

where $k$ is the wave number of the sodium laser. $\mu_m^+(H)$ is the upper moment, which can be expressed as:

$$\mu_m^+(H) = \int_H^\infty dz \cdot C_N^2(z) \cdot z^m \quad (16)$$

$\mu_m^-(H)$ is the lower moment, which can be expressed as:

$$\mu_m^-(H) = \int_0^H dz \cdot C_N^2(z) \cdot z^m \quad (17)$$

The cone effect is characterized by the parameter $d_0$. The value of $d_0$ is related to the altitude of the LGS, the profile of the turbulence and the observing wavelength. Although the relatively low altitude of the Rayleigh LGS produces inadequate sampling of atmospheric turbulence, more recent measurements have shown that 60% of the turbulence is mainly located within the first two kilometers above ground. In table 2 we presented the calculated values of $d_0$ and the corresponding $\sigma_{cone}^2$, assuming the Rayleigh LGS is located near 9 *km* altitude with a wavelength of 589 *nm* for zenith angels of 0° and 45° with Hunfnagel [23], Mauna Kea background and Mauna Kea average [24] turbulence models.

Table 2. Values of $d_0$ and cone effect under three typical turbulence models

| Turbulence model | Zenith Angle (°) | $d_0$@589nm (m) | $\sigma_{cone}^2$ (rad²) |
| --- | --- | --- | --- |
| Hunfnagel | 0° | 0.682 | 0.25 |
| | 45° | 0.56 | 0.35 |
| Mauna Kea background | 0° | 1.62 | 0.06 |
| | 45° | 1.31 | 0.08 |
| Mauna Kea average | 0° | 1.06 | 0.12 |
| | 45° | 0.86 | 0.17 |

As shown in this table, for all three profiles, the wavefront errors induced by cone effect are much less than 1 rad² which corresponding to a Strehl ratio to 0.37 [3]. In light of this, the cone effect of the Rayleigh LGS used for pre-correction can be neglected.

### 4.3 The launching and receiving time sequencing of multiple pulses

The receiving photon return of Rayleigh plume exists continuously all along the up-path while the laser was projected from the LLT. The turbulence detected by the Rayleigh plume at different altitude will be different. Thus, the photon return of the multiple pulses at different altitude should be separated, which is related to the launching and receiving time sequencing of the multiple pulses [25].

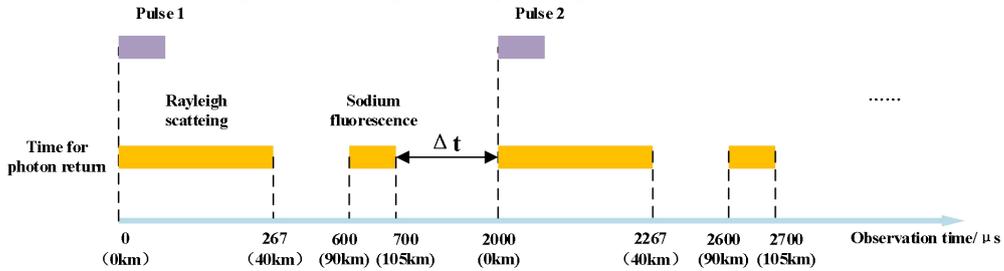

Fig. 10. Timeline for the pulse in the uplink path and the photon return the LLT can be received. The purple boxes indicate the time of the pulse projected from the LLT, the pulse width was neglected. The Rayleigh scatting was assumed disappeared with the altitude exceed 40 *km*, the sodium layer was assumed range from 90 to 105 *km*.

The timeline for the Rayleigh photons received by the LLT was shown in Fig. 10. The photon return for altitude up to 40 *km* is assumed to be disappeared as discussed previously. The photons can be detected lasts 0 - 267 *μs* after the first pulse was projected. The altitude rang of sodium layer was assumed to be from 90 *km* to 105 *km*. The photon return of the sodium LGS can be detected lasts 600-700 *μs*. Considering the altitude angle up to 30°, the uplink path is nearly 200 *km*. The time interval between the adjacent pulses is nearly 2000 *μs*, and the corresponding repetition frequency is 500 Hz. With the increase of the altitude angle, the time interval Δt can be decreased, and the laser can be pulsed at a higher repetition rate. For the LLT pointing to the zenith and Δt is 0, the critical condition occurs that Rayleigh photon return of pulse 2 interfered with the sodium photon return of pulse 1, which means the turbulence detected by the two kinds of return light will be mingled, the maximum repetition frequency occurs around 1500 Hz.

## 5. Conclusions

In this study, we experimentally verified the MSISE-90 atmospheric mode is applicable for predicting the photon return of Rayleigh plume. We then analyzed the Rayleigh plume used as a "by-product" of sodium LGS to detect the ground layer turbulence on the uplink path for pre-correction of the sodium LGS. Corresponding results can be summarized as follows:

(1) The MSISE-90 atmospheric mode can be used for predicting the photon return of "by-product" Rayleigh plume of the uplink sodium laser, the largest difference between the theoretical and experimental results is nearly 20% which is caused mainly by the Mie scattering from dust and aerosols.

(2) The measurement error depends on SNR and the spot size of the "by-product" Rayleigh LGS. For a fixed SNR of 10, the key parameter is the center altitude of the range gate which determines the range gate depth and spot elongation. For $r_0$ is 5 *cm* or 10 *cm*, the center altitude of range gate can be selected as 9 *km*. The corresponding value of range gate depth are 3 *km*, 1.1 *km*, 0.56 *km* and 0.28 *km* for subaperture size of 3 *cm*, 5 *cm*, 7 *cm* and 10 *cm*.

(3) The turbulence detected by the sodium fluorescence or the Rayleigh scattering of the previous pulse may be mingled with the Rayleigh scattering of the next pulse, which can be avoided by adjusting the interval between adjacent pulses. In this case, the laser can be pulsed at the highest rate of 1500 Hz.

In the next step, an experimental verification of wavefront sensing for the turbulence on the uplink path by the "by-product" Rayleigh plume will be conducted.

## Acknowledgments

We are thankful for the advice, guidance and support from Doctor Keran Deng from the Key Laboratory on Adaptive Optics, Chinese Academy of Sciences.

## Funding

This work was supported by National Natural Science Foundation of China (Grant No.6210032680), The Natural Science Foundation of Chongqing, China (Grant No. cstc2020jcyj-msxmX0727), Scientific Research Project of Chongqing Technology and Business University (Grant NO.1952016), Chongqing Key Laboratory of Manufacturing Equipment Mechanism Design and Control, China (Grant NO.KFJJ2019074),